\documentclass[%
 preprint,prb,
 amsmath,amssymb,
 aps,superscriptaddress
]{revtex4-1}

\usepackage{graphicx}
\usepackage{dcolumn}
\usepackage{bm}
\usepackage{subfig}
\usepackage{tabularx}
\usepackage{booktabs}
\usepackage{color}
\usepackage[normalem]{ulem}
\usepackage{natbib}

\begin{document}

\preprint{APS/123-QED}

\title{Benchmarking Non-perturbative Many-Body Approaches in the Exactly Solvable Hatsugai-Kohmoto Model}

\author{Hui Li}
 \email{physicslihui@zju.edu.cn}
\affiliation{%
Institute for Advanced Study in Physics and School of Physics, Zhejiang University, Hangzhou 310027, China
} 

\author{Ziyu Li}%
\affiliation{%
School of Physics, Peking University, Beijing 100871, China
}

\author{Run-yu Chen}
\affiliation{%
Institute for Advanced Study in Physics and School of Physics, Zhejiang University, Hangzhou 310027, China
}

\date{\today}

\begin{abstract}
The accurate simulation of strongly correlated electron systems remains a central challenge in condensed matter physics, motivating the development of various non-perturbative many-body methods. Such methods are typically benchmarked against the numerical exact determinant quantum Monte Carlo (DQMC) in the Hubbard model; however, DQMC is limited by the fermionic sign problem and the uncertainties of numerical analytic continuation. To address these issues, we use the exactly solvable Hatsugai–Kohmoto (HK) model as a benchmarking platform to evaluate three many-body approximations: $GW$, $HGW$, and $SGW$. We compare the Green’s functions, spectral functions, and response functions obtained from these approximations with the exact solutions. Our analysis shows that the $GW$ approximation, often considered insufficient for describing strong correlation, exhibits a previously unreported solution branch that accurately reproduces Mott physics in the HK model. In addition, using a covariant formalism, we find that $HGW$ provides an accurate description of charge response, while $SGW$ performs well for spin correlations. Overall, our work demonstrates that the HK model can effectively benchmark many-body approximations and helps refine the understanding of $GW$ methods in strongly correlated regimes.
\end{abstract}

\maketitle


\section{Introduction}
The calculation of single-particle quantities and response functions in correlated electronic systems is a long-standing challenge and one of the central problems in condensed matter physics. Computational methods based on single-particle pictures, such as Hartree-Fock mean-field theory, perturbative treatments, or DFT often cannot describe these systems accurately enough\cite{mott1949basis,imada1998metal,DFT,keimer2015quantum,WXG2006_RMP}, especially when strongly correlated effects dominate. Consequently, a wide range of many-body approximation methods\cite{senechal2004theoretical}, such as FLEX\cite{bickers1989FLEX}, TPSC\cite{vilk1997TPSC,tremblay2011two}, DMFT, and its cluster extensions\cite{rohringer2018diagrammatic,kotliar2001cellular,toschi2007dynamical,vuvcivcevic2017trilex,maier2005quantum}, have been developed. These methods are frequently benchmarked against the numerically exact determinant quantum Monte Carlo (DQMC) method in the context of the Hubbard model\cite{scalapino1981dqmc,blankenbecler1981dqmc,schafer2021tracking}. The Hubbard model is widely regarded as the simplest lattice model of the strongly correlated fermions and serves as a fundamental platform for studying phenomena like Mott physics and high-temperature superconductivity in cuprates\cite{hubbard1963electron,WXG2006_RMP,qin2022hubbard}.

Among these approaches, a large class of methods is constructed by truncating the infinite hierarchy structure of Dyson–Schwinger equations, playing an essential role in theoretical models and first-principles materials calculations\cite{GW_review_2019,Hedin,kadanoff1961theory,sun2021modified,cubic2018,Fan2020,xiong2025,kutepov2017one}. These approaches can be broadly categorized into two types: those that truncate high-order vertex functions, such as $GW$\cite{GW_review_2019,Hedin,Aryasetiawan2008}, and those that truncate high-order correlation functions, such as $HGW$\cite{sun2021modified} (also referred to as $G_0G$\cite{kadanoff1961theory}), cubic\cite{cubic2018}, and quartic theories\cite{Fan2020}. To evaluate two-body correlation functions within different approximations, we have developed a covariant formalism that defines them via the system's response to the external perturbation \cite {cGW_2023}. This framework uniquely determines response functions for a given approximation and automatically preserves the Ward–Takahashi identity (WTI)\cite{peskin2018introduction} and fluctuation-dissipation theorem (FDT)\cite{altland_simons_2010}. It has shown good agreement in benchmarks of both spin and charge correlations in the two-dimensional Hubbard model. Comparative studies with DQMC further clarify the distinct characteristics of different approximations: vertex-based methods, such as $GW$, become less accurate at stronger interactions and fail to capture Mott physics\cite{GW_review_2019}. By contrast, the $HGW$ method correctly describes the Mott gap, although it tends to overestimate the gap width\cite{sun2021modified}. Meanwhile, the $GW$ approximation in the spin-channel offers improved accuracy in calculating spin correlations\cite{cGW_2023}.

However, benchmarking with DQMC in the Hubbard model faces two major challenges: (1) the high computational cost limits simulations to relatively small system sizes, and when the fermion sign problem is present, it becomes difficult to obtain reliable results at low temperatures or in large systems\cite{pan2022sign,becca2017qmc}; (2) spectral functions obtained from DQMC depends on numerical analytic continuation, such as the maximum entropy method\cite{maxent2016,jarrell1996bayesian}, which introduces additional uncertainties and reduces the reliability of such benchmarks.

The Hatsugai-Kohmoto (HK) model has recently attracted renewed interest as a system with infinite-range interactions\cite{hatsugai1992hk,hatsugai1996hk,zhongyin2023,zhongyin2025review,phillips2020exact}. It provides an exactly solvable example of non-Fermi liquid behavior and featureless Mott insulating phases in arbitrary spatial dimensions. The solvability stems from the model's locality in momentum space, which allows the Hamiltonian to be diagonalized independently at each momentum point. Owing to its closed set of equations of motion, the Green's functions can be solved exactly, allowing spectral functions to be obtained analytically without numerical continuation. These properties make the HK model particularly suitable for benchmarking many-body approximation methods and studying their capability to capture the strongly correlated phenomenon, especially Mott physics.

In this article, we apply three non-perturbative many-body approximate theories, including $HGW$, $GW$, and spin $GW$ ($SGW$), which represent different truncation strategies, to the HK model and examine their corresponding response functions using the covariant formalism. Consistent with findings in the Hubbard model, $HGW$ demonstrates the ability to describe the Mott gap and proves effective in capturing charge correlations, while $SGW$ remains accurate in representing spin correlations. More notably, beyond conventional understanding, we find that $GW$ exhibits a nontrivial Mott solution that accurately captures the essential Mott characteristics of the HK model.

The remaining part of this article is organized as follows. In Sec.~\ref{Sec:Theoretical}, we introduce the HK model and its exact solutions, followed by the formulation of the $GW$, $HGW$, and $SGW$ approximations for the Green’s function, along with their covariant response theories. Sec.~\ref{Sec:Numerical} presents a numerical comparison of the Green’s functions and spectral functions obtained from $GW$, $HGW$, and $SGW$ against exact results, and examines the charge and spin response functions in momentum space. The conclusion and discussion are given in Sec.~\ref{Sec:Conclusion}.

\section{Theorectical Formalism}\label{Sec:Theoretical}
\subsection{HK Model}
The Hamiltonian of the HK model in a two-dimensional square lattice can be expressed as:
\begin{equation}
\hat H=\sum_{\vec k\sigma}(\varepsilon_{\vec k}-\mu)\hat{c}_{{\vec k}\sigma}^\dagger \hat{c}_{{\vec k}\sigma}+U\sum_{\vec k} \hat n_{{\vec k}\uparrow}\hat n_{{\vec k}\downarrow},
\end{equation}
where $\hat c_{\vec k\sigma}^\dagger$ creates an electron at momentum $\vec k=(k_x,k_y)$ with spin $\sigma$, and $\hat n_{\vec{k}\sigma}=\hat c_{\vec k\sigma}^\dagger \hat c_{\vec k\sigma}$. For a $L\times L$ lattice, the momentum takes the values as $\vec k=\frac{2\pi}{L}(m_x,m_y)$, $m_x,m_y=-\frac{L}{2},-\frac{L}{2}+1,\cdots,\frac{L}{2}-1$. The non-interacting electron dispersion is $\varepsilon_{\vec k}=-2t(\cos k_x+\cos k_y)$, with nearest-neighbor-hopping $t$. $\mu$ is the chemical potential and $U$ is the strength of the HK interaction.

A key feature of the HK model is that the interaction is local in momentum space, allowing the Hamiltonian to be decoupled into different momentum subspace as $H=\sum_{\vec k}\hat H_{\vec k}$, with
\begin{equation}
    \hat H_{\vec k}=\sum_{\sigma}(\varepsilon_{\vec k}-\mu)\hat{c}_{{\vec k}\sigma}^\dagger \hat{c}_{{\vec k}\sigma}+U\hat n_{{\vec k}\uparrow}\hat n_{{\vec k}\downarrow}.
\end{equation}
Each momentum subspace is only 4-dimensional, with basis as $\left|  0\right>_{\vec k}$, $\left| \uparrow \right>_{\vec k}=\hat c_{\vec k\uparrow}^\dagger\left|  0\right>_{\vec k}$, $\left| \downarrow \right>_{\vec k}=\hat c_{\vec k\downarrow}^\dagger\left|  0\right>_{\vec k}$ and $\left| \uparrow\downarrow \right>_{\vec k}=\hat c_{\vec k\uparrow}^\dagger \hat c_{\vec k\downarrow}^\dagger\left|  0\right>_{\vec k}$, and can thus be solved exactly. The momentum-resolved particle number is given by::
\begin{equation}
    n_{\vec k\sigma}=\frac{f_F(\varepsilon_{\vec k}-\mu)}{f_F(\varepsilon_{\vec k}-\mu)+1-f_F(\varepsilon_{\vec k}-\mu+U)}.
\end{equation}
Here $f_F(x)=1/(e^{\beta x}+1)$ is the Fermi-Dirac distribution function, and $\beta=1/T$ is the inverse temperature.

In this article, we consider two kinds of response function: the charge susceptibility $\chi_c$ and the spin susceptibility $\chi_s$. The charge susceptibility can be simulated through 
\begin{equation}
    \chi_c=\frac{\partial n}{\partial \mu}=\frac{1}{L^2}\sum_{\vec k\sigma}\frac{\partial n_{\vec k\sigma}}{\partial\mu}.
\end{equation}
To calculate the spin susceptibility, one needs to add the Zeeman coupling term to the original Hamiltonian,
\begin{equation}
    \hat H_Z=-h\sum_{\vec k\sigma}\sigma\hat n_{\vec k\sigma},
\end{equation}
which leads to a spin-dependent dispersion $\varepsilon_{\vec k\sigma}(h)=\varepsilon_{\vec k}-h\sigma$ and the polarized particle distribution:
\begin{equation}
    n_{\vec k\sigma}(h)=\frac{e^{-\beta(\varepsilon_{\vec k\sigma}(h)-\mu)}+e^{-\beta(2\varepsilon_{\vec k}-2\mu+U)}}{1+e^{-\beta(\varepsilon_{\vec k\uparrow}(h)-\mu)}+e^{-\beta(\varepsilon_{\vec k\downarrow}(h)-\mu)}+e^{-\beta(2\varepsilon_{\vec k}-2\mu+U)}}.
\end{equation}
The magnetization is $M(h)=\frac{1}{L^2}\sum_{\vec k}(n_{\vec k\uparrow}-n_{\vec k\downarrow})$, and the corresponding spin susceptibility can be obtained:
\begin{equation}
    \chi_s=\frac{\partial M}{\partial h}\left.\right|_{h\rightarrow0}.
\end{equation}

To study the single-particle properties, one should calculate the one-body Green's function. To simplify, we take the coherent state path integral with the Matsubara action as our starting point,
\begin{align}
    &Z=\int D[\psi^*\psi]e^{-S[\psi^*\psi]},\nonumber\\
    &S[\psi^*\psi]=\sum_{\vec k}\sum_{\sigma}\int d\tau[\psi^*_{\vec k\sigma}(\tau)(\partial_\tau+\varepsilon_{\vec k}-\mu)\psi_{\vec k\sigma}(\tau)+\frac{U}{2}\psi_{\vec k\sigma}^*(\tau)\psi_{\vec k\sigma}(\tau)\psi_{\vec k\bar\sigma}^*(\tau)\psi_{\vec k\bar\sigma}(\tau)],
\end{align}
where $0<\tau<\beta$, and $\bar\sigma$ refers to the opposite spin of $\sigma$. The Green's function is defined as
\begin{equation}
    G_{\vec k\sigma}(1,2)=\left< \psi^*_{\vec k\sigma}(2)\psi_{\vec k\sigma}(1) \right>,
\end{equation}
where the notation $1=\tau_1$ labels the imaginary index, and the average is defined by
\begin{equation}
    \left< \cdots \right>=\frac{1}{Z}\int D[\psi^*\psi]\cdots e^{-S[\psi^*\psi]}.
\end{equation}

This Green's function can be exactly solved and takes the expression:
\begin{equation}
    G_\sigma(\vec k,i\omega_n)=\frac{1-n_{\vec k\bar\sigma}}{i\omega_n-(\varepsilon_{\vec k}-\mu)}+\frac{n_{\vec k\bar\sigma}}{i\omega_n-(\varepsilon_{\vec k}-\mu+U)}.
\end{equation}
Here $i\omega_n=\frac{(2n+1)\pi}{\beta}$ is the Matsubara frequency. The corresponding imaginary-time Green's function for $0 < \tau < \beta$ is:
\begin{align}
G_\sigma(\vec k,\tau)=&-(1-n_{\vec k\bar\sigma})e^{-(\varepsilon_{\vec k}-\mu)\tau}[1-f(\varepsilon_{\vec k}-\mu)]\nonumber\\
&-n_{\vec k\bar\sigma}e^{-(\varepsilon_{\vec k}-\mu+U)\tau}[1-f(\varepsilon_{\vec k}-\mu+U)].
\end{align}
The corresponding retarded Green's function is obtained through the analytic continuation $i\omega_n \to \omega + i\eta$, 
\begin{equation}
    G_\sigma(\vec k,\omega)=   \frac{1-n_{\vec k\bar\sigma}}{\omega-(\varepsilon_{\vec k}-\mu)+i\eta}+\frac{n_{\vec k\bar\sigma}}{\omega-(\varepsilon_{\vec k}-\mu+U)+i\eta}.
\end{equation}
where we set $\eta = 0.1$ for numerical convenience.
The spectral function, therefore, is
\begin{equation}
    A_{\sigma}(\vec k,\omega)=\frac{1}{\pi}\mathrm{Im}G_{\sigma}(\vec k,\omega)=(1-n_{\vec k\bar\sigma})\delta(\omega-(\varepsilon_{\vec k}-\mu))+n_{\vec k\bar\sigma}\delta(\omega-(\varepsilon_{\vec k}-\mu+U)),\label{eq:exact_spectral}
\end{equation}
which has two peaks at $\omega=\varepsilon_{\vec k}-\mu$ and $\omega=\varepsilon_{\vec k}-\mu+U$. Consequently, the HK model exhibits a Mott gap given by $\Delta_{\mathrm{Mott}} = U$.

The exact solvability of the HK model for key properties, including Green’s functions, spectral functions, and response functions, makes the HK model an ideal benchmark system. In the following sections, we use these rigorous results to benchmark the performance of $GW$, $HGW$, and $SGW$.

\subsection{Dyson-Schwinger Equation}
In the path-integral formulation, the Matsubara action can be written as
\begin{equation}
    S=-\sum_{\vec k}\int d(\bar1\bar2) \psi^*_{\vec k}(\bar1)(G^{-1}_0)_{\vec k}(\bar1,\bar2)\psi_{\vec k}(\bar2)+\frac{1}{2}\sum_{\vec k}\int d(\bar1\bar2)\psi^*_{\vec k}(\bar1)\psi_{\vec k}(\bar1) V(\bar1,\bar2)\psi^*_{\vec k}(\bar2)\psi_{\vec k} (\bar2).  
\end{equation}
Here, we use the notation $\bar 1=(\tau_1,\sigma_1)$, and $\int d\bar 1=\sum_{\sigma_1}\int_0^\beta d\tau_1$. The inverse free propagator and the interaction function take the form
\begin{equation}
    (G^{-1}_0)_{\vec k}(\bar1,\bar2)=(\partial_{\tau_1}+\varepsilon_{\vec k}-\mu)\delta_{\sigma_1\sigma_2}\delta(\tau_1,\tau_2),
\end{equation}
\begin{equation}
    V(\bar1,\bar2)=U\delta(\tau_1,\tau_2)\delta_{\sigma_1\bar\sigma_2},
\end{equation}
where $\delta(1,2)$ is the Dirac delta function, and $\delta_{\sigma_1\sigma_2}$ is the Kronecker symbol.

To derive the Dyson-Schwinger equations (DSE), we add the source term to the action:
\begin{equation}
    S[\psi^*,\psi;J]=S[\psi^*,\psi]-\sum_{\vec k}\int d\bar 1 J_{\vec k}(\bar 1)\psi^*_{\vec k}(\bar 1)\psi_{\vec k}(\bar 1).
\end{equation}
Given that the functional integration measure is invariant under an infinitesimal translation of the field $\psi$, we have
\begin{equation}
\int \mathcal{D}[\psi^*\psi] \frac{\delta}{\delta \psi_{\vec k}(\bar{1})} \{ \psi_{\vec k}^*(\bar{2}) e^{-S[\psi^*,\psi;J]} \} = 0.\label{eq:DS}
\end{equation}
This identity directly leads to the Dyson-Schwinger equation (DSE), which establishes a relation between the Green's function and its functional derivative:
\begin{align}
\delta(\bar{1},\bar{2}) - \int d\bar{3}H^{-1}_{\vec{k}}(\bar{1},\bar{3}) G_{\vec{k}}(\bar{3},\bar{2}) + \int d\bar{3}V(\bar{1},\bar{3}) \frac{\delta G_{\vec{k}}(\bar{1},\bar{2})}{\delta J_{\vec{k}}(\bar{3})} = 0.\label{eq:DS-1}
\end{align}
The Hartree propagator H is defined by
\begin{equation}
H^{-1}_{\vec k}(\bar1,\bar2)=(G^{-1}_0)_{\vec k}(\bar1,\bar2)+\delta(\bar1,\bar2)v_{\vec k}(\bar1),
\end{equation}
and the effective potential is
\begin{equation}
    v_{\vec k}(\bar 1)=J_{\vec k}(\bar1)-\int d\bar 3V(\bar1,\bar3)n_{\vec k}(\bar3).
\end{equation}
Different many-body methods employ distinct strategies for approximating the functional derivative $\frac{\delta G}{\delta J}$. This work focuses on two such approximations: vertex truncation and cumulant truncation, which yield the $GW$ and $HGW$ methods, respectively.

\subsection{$GW$, $HGW$ and spin-$GW$ approximations}
As for the $GW$ approximation, we define Hedin's vertex function as
\begin{equation}
    \Lambda_{\vec k}(\bar1,\bar2,\bar3)\equiv\frac{\delta G_{\vec k}^{-1}(\bar1,\bar2)}{\delta v_{\vec k}(\bar3)},\label{eq:GW-vertex}
\end{equation}
and the screened potential is defined as
\begin{equation}
    W_{\vec{k}}(\bar1,\bar2)\equiv\int d(\bar 3)\frac{\delta v_{\vec{k}}(\bar1)}{\delta J_{\vec{k}}(\bar3)}V(\bar2,\bar3).\label{eq:GW-W}
\end{equation}
Therefore, the functional derivative term can be written as $\frac{\delta G}{\delta J}=-G\Lambda G$. Then one can obtain the famous Hedin's equation by substituting Eqs.~(\ref{eq:GW-vertex},\ref{eq:GW-W}) to Eq.(\ref{eq:DS-1}), and setting $J\rightarrow 0$:
\begin{subequations}
\label{eq:Hedin-system}
\begin{align}
    &G^{-1}_{\vec k}(\bar1,\bar2)=H^{-1}_{\vec k}(\bar1,\bar2)-\Sigma_{\vec k}(\bar1,\bar2),\label{eq:GW-Dyson}\\
    &\Sigma_{\vec k}(\bar1,\bar2)=-\int d(\bar4\bar5)G_{\vec k}(\bar1,\bar4)\Lambda_{\vec k}(\bar4,\bar2,\bar5)W_{\vec k}(\bar5,\bar1),\label{eq:Hedin-Sigma}\\
    &W^{-1}_{\vec k}(\bar1,\bar2)=V^{-1}(\bar1,\bar2)-\Pi_{\vec k}(\bar1,\bar2),\label{eq:GW-screened}\\
    &\Pi_{\vec k}(\bar1,\bar2)=\int d(\bar3\bar4)G_{\vec k}(\bar1,\bar3)\Lambda_{\vec k}(\bar3,\bar4,\bar2)G_{\vec k}(\bar4,\bar1),\label{eq:Hedin-P}\\
    &H^{-1}_{\vec k}(\bar1,\bar2)=(G^{-1}_0)_{\vec k}(\bar1,\bar2)-\delta(\bar1,\bar2)\int d\bar 3V(\bar1,\bar3)G_{\vec k}(\bar3,\bar3)\label{eq:GW-H}.
\end{align}
\end{subequations}
Here, $\Pi$ is the polarization function, and $\Sigma$ is the self-energy. The full Hedin's equations cannot be solved exactly, which requires a truncation of the vertex. The $GW$ approximation is defined by the lowest-order truncation,
\begin{equation}
    \Lambda_{\vec k}(\bar1,\bar2,\bar3)\approx\frac{\delta H^{-1}_{\vec k}(\bar1,\bar2)}{\delta v_{\vec k}(\bar3)}=\delta(\bar1,\bar2)\delta(\bar1,\bar3),
\end{equation}
which leads to the $GW$ approximation with the self-energy and the polarization:
\begin{equation}
    \Sigma_{\vec k}(\bar1,\bar2)=-G_{\vec k}(\bar1,\bar2)W_{\vec k}(\bar2,\bar1),\label{eq:GW-Sigma}
\end{equation}
\begin{equation}
    \Pi_{\vec k}(\bar1,\bar2)=G_{\vec k}(\bar1,\bar2)G_{\vec k}(\bar2,\bar1).\label{eq:eq:GW-P}
\end{equation}
These equations can be solved self-consistently.

Beyond vertex truncation, a fundamentally different strategy is to close the equations by directly truncating the hierarchy of correlation functions. Taking the functional derivative of Eq.~\ref{eq:DS-1} with respect to the source $J$ leads to:
\begin{align}
0 = &\int d\bar{3}\frac{\delta H^{-1}_{\vec k}(\bar{1},\bar{3})}{\delta J_{\vec k}(\bar{4})} G_{\vec k}(\bar{3},\bar{2}) + \int d\bar{3}H^{-1}_{\vec k}(\bar{1},\bar{3}) \frac{\delta G_{\vec k}(\bar{3},\bar{2})}{\delta J_{\vec k}(\bar{4})} 
- \int d\bar{3} V(\bar{1},\bar{3}) \frac{\delta^2 G_{\vec k}(\bar{1},\bar{2})}{\delta J_{\vec k}(\bar{4}) \delta J_{\vec k}(\bar{3})},
\label{eq:DS-2}
\end{align}
which connects the one-body Green's function $G$, the two-body correlator $\delta G / \delta J$, and the three-body correlator $\delta^2 G / \delta J^2$. By repeatedly applying functional derivatives, one can derive an infinite hierarchy of Dyson–Schwinger equations (DSEs). To make these equations solvable, the correlation functions must be truncated at a certain order.

The simplest truncation, $\delta G / \delta J \to 0$ applied to Eq.~\eqref{eq:DS-1}, leads to the Hartree approximation, equivalent in this case to the Hartree–Fock approximation:
\begin{equation}
G_{\mathrm{HF}}(\bar{1},\bar{2}) = H(\bar{1},\bar{2}).
\end{equation}

Beyond this mean-field level, a natural approximation at the lowest order is achieved by setting $\delta^2 G / \delta J^2 \to 0$ in Eq.~\eqref{eq:DS-2}, which leads to the $HGW$ equations:
\begin{subequations}
\label{eq:HGW-system}
\begin{align}
&G^{-1}_{\vec{k}}(\bar{1},\bar{2}) = H^{-1}_{\vec k}(\bar{1},\bar{2}) - \Sigma_{\vec k}(\bar{1},\bar{2}), \label{eq:HGW-Dyson} \\
&\Sigma_{\vec k}(\bar{1},\bar{2}) = -H_{\vec k}(\bar{1},\bar{2}) W_{\vec k}(\bar{2},\bar{1}), \label{eq:HGW-Sigma} \\
&W^{-1}_{\vec k}(\bar{1},\bar{2}) = V^{-1}(\bar{1},\bar{2}) - \Pi_{\vec k}(\bar{1},\bar{2}), \label{eq:HGW-screened} \\
&\Pi_{\vec k}(\bar{1},\bar{2}) = H_{\vec k}(\bar{1},\bar{2}) G_{\vec k}(\bar{2},\bar{1}), \label{eq:HGW-P} \\
&H^{-1}_{\vec k}(\bar{1},\bar{2}) = (G^{-1}_0)_{\vec{k}}(\bar{1},\bar{2}) - \delta(\bar{1},\bar{2}) \int d\bar{3} V(\bar{1},\bar{3}) G_{\vec{k}}(\bar{3},\bar 3). \label{eq:HGW-H}
\end{align}
\end{subequations}

Comparing the $GW$ and $HGW$ sets of equations reveals that the latter can be obtained by replacing certain instances of the full Green's function $G$ in the $GW$ formalism with the Hartree propagator $H$.

Different formulations of the interaction, while mathematically equivalent, can inspire distinct many-body approximations. Beyond the charge channel addressed above, the interaction can be expressed in the spin channel to specifically target magnetic correlations, providing access to the $SGW$ method. Here, we express the HK interaction in the form\cite{Aryasetiawan2008,vuvcivcevic2017trilex}:
\begin{equation}
    U\sum_{\vec k}n_{\vec{k}\uparrow} n_{\vec{k}\downarrow}=-\frac{U}{6}\sum_{\vec k}\sum_{a=x,y,z} S_{\vec{k}}^a S_{\vec{k}}^a,
\end{equation}
where $S_{\vec k}^a=\sum_{\sigma_1\sigma_2}\psi^*_{\vec k\sigma_1}\tau^a_{\sigma_1\sigma_2}\psi_{\vec{k}\sigma_2}$ is the spin operator with $\tau^a$ the Pauli matrices. Without loss of generality, we rewrite the action into the following form:
\begin{align}
    S[\psi^*,\psi]=&-\sum_{\vec k}\sum_{\sigma_1\sigma_2}\int d(12) \psi^*_{{\vec k}\sigma_1}(1)(G^{-1}_0)_{{\vec k};\sigma_1,\sigma_2}(1,2)\psi_{{\vec k}}(2)\nonumber\\
    &+\frac{1}{2}\sum_{\vec k}\sum_{ab}\int d(12)S^a_{\vec k}(1)V^{ab}(1,2)S^b_{\vec k}(2).  
\end{align}
Here, the interaction matrix is given by $V^{ab}(1,2)=-\frac{U}{3}\delta_{ab}\delta(1,2)$. To derive the generalized $GW$ equations, we introduce a source term coupled to the spin operator:
\begin{equation}
    S[\psi^*,\psi;\vec J]=S[\psi^*,\psi]-\sum_{\vec{k}}\sum_a\int d(1)J^a_{\vec k}(1)S^a_{\vec k}(1).
\end{equation}
Given the spin structure of the interaction, it is useful to represent quantities as matrices in spin space
\begin{equation}
    \underline{X}=
    \begin{bmatrix}
X_{\uparrow\uparrow} & X_{\uparrow\downarrow}\\
X_{\downarrow\uparrow} & X_{\downarrow\downarrow}
    \end{bmatrix},
\end{equation}
with the trace defined as $\mathrm{tr}[\underline{X}]=X_{\uparrow\uparrow}+X_{\downarrow\downarrow}$. Following the analogous procedure for deriving Eq.~(\ref{eq:DS}) then leads to the Dyson-Schwinger equation,
\begin{equation}
 \delta(1,2)\underline{I}=\int d(3)\underline{H}^{-1}_{\vec k}(1,3)\underline{G}_{\vec k}(3,2)-\sum_{ab}\int d(3)\underline{\tau}^a V^{ab}(1,3)\frac{\delta \underline{G}_{\vec k}(1,2)}{\delta J^b_{\vec k}(3)},   \label{eq:SGW-DSE}
\end{equation}
where $\underline{I}$ is the identity matrix in spin space. The Hartree propagator is defined as,
\begin{equation}
    \underline{H}^{-1}_{\vec k}(1,2)=\underline{G}^{-1}_{0\vec k}(1,2)+\delta(1,2)\sum_av^a_{\vec k}(1)\underline{\tau}^a,
\end{equation}
and the effective potential is
\begin{equation}
    v^a_{\vec k}(1)\equiv J^a_{\vec k}-\sum_b\int d(3)V^{ab}(1,3)\mathrm{tr}[\underline{G}_{\vec k}(3,3)\underline{\tau}^b].
\end{equation}
To describe the spin interaction channel, we define Hedin's vertex as
\begin{equation}
    \underline\Lambda^a_{\vec k}(1,2,3)\equiv\frac{\delta \underline G^{-1}_{\vec k}(1,2)}{\delta v^a_{\vec k}(3)},\label{eq:SGW-vertex}
\end{equation}
and the screened potential
\begin{equation}
    W^{ab}_{\vec k}(1,2)\equiv \sum_c\int d(3)\frac{\delta v^a_{\vec k}(1)}{\delta J^c_{\vec k}(3)}V^{bc}(2,3).\label{eq:SGW-W}
\end{equation}
Substituting Eqs.~(\ref{eq:SGW-vertex}) and (\ref{eq:SGW-W}) into the Eq.~(\ref{eq:SGW-DSE}) yields the generalized Hedin equations:
\begin{subequations}
\label{eq:SGW-system}
\begin{align}
    &\underline{G}^{-1}_{\vec k}(1,2)=\underline{H}^{-1}_{\vec k}(1,2)-\underline\Sigma_{\vec k}(1,2),\label{eq:SGW-dyson}\\
    &\underline\Sigma_{\vec k}(1,2)=-\sum_{ab}\int d(34)\underline{\tau}^a\underline{G}_{\vec k}(1,4)\underline\Lambda^b_{\vec k}(4,2,3)W^{ba}_{\vec k}(3,1),\\
    &(W^{-1})^{ab}_{\vec k}(1,2)=(V^{-1})^{ab}-\Pi^{ab}_{\vec k}(1,2),\\
    &\Pi^{ab}_{\vec k}(1,2)=\int d(34)\mathrm{Tr}[\underline{\tau}^a\underline{G}_{\vec k}(1,3)\underline{\Lambda}^b_{\vec k}(3,4,2)\underline{G}_{\vec k}(4,1)],\\
    &\underline{H}^{-1}_{\vec k}(1,2)=\underline{G}^{-1}_{0\vec k}(1,2)-\delta(1,2)\sum_{ab}\int d(3)\underline{\tau}^aV^{ab}(1,3)\mathrm{tr}[\underline{G}_{\vec k}(3,3)\underline{\tau}^b].\label{eq:SGW-H}
\end{align}
\end{subequations}
The $SGW$ approximation is obtained by applying the corresponding lowest-order vertex approximation, $\underline{\Lambda}^a_{\vec k}(1,2,3)\approx\underline{\tau}^a\delta(1,2)\delta(1,3)$. The corresponding self-energy and polarization function are:
\begin{equation}
    \underline\Sigma_{\vec k}(1,2)=-\sum_{ab}\underline{\tau}^a\underline{G}_{\vec k}(1,2)\underline\tau^bW^{ba}_{\vec k}(2,1),\label{eq:SGW-sigma}
\end{equation}
\begin{equation}
\Pi^{ab}_{\vec k}(1,2)=\mathrm{Tr}[\underline{\tau}^a\underline{G}_{\vec k}(1,2)\underline{\tau}^b\underline{G}_{\vec k}(2,1)].\label{eq:SGW-P}
\end{equation}
As this formalism specifically incorporates the spin channel, we refer to it as the $SGW$ approximation.

We have thus formulated three non-perturbative methods ($GW$, $HGW$, and $SGW$) based on distinct truncation schemes for the HK model, all extending beyond mean-field theory. These approaches, which share comparable numerical complexity, can be self-consistently solved to determine the one-body Green's function $G$ and the screened potential $W$, thereby establishing a unified beyond-mean-field landscape.

\subsection{Covariant Framework For two-body correlation functions}

The HK model is characterized by two key response functions: the charge susceptibility and the spin susceptibility. In Ref~[\cite{cGW_2023}], we introduced the covariant framework for computing two-body correlation functions. This framework automatically preserves both the fluctuation-dissipation theorem (FDT) and the Ward-Takahashi identity (WTI). Within this scheme, a generic two-body correlation function $\chi_{XY}(1,2) = \langle X(1) Y(2) \rangle$ is defined as the linear response of the observable $\langle X \rangle$ to an external field coupled to $Y$, where $X$ and $Y$ are binary composite operators.

The general computational procedure is formulated as follows. First, the corresponding source term is introduced into the action, $S[\psi^*,\psi;\phi] = S[\psi^*,\psi]-\sum_{\vec k}\int d(1)\ \phi_{Y\vec k}(1) Y_{\vec k}(1)$ and the correlation can be obtained by $\chi_{XY}(1,2)=\delta \left< X_{\vec k}(1) \right>/\delta \phi_{Y\vec k}(2)$. Next, the off-shell $GW$, $HGW$, or $SGW$ equations (with the source $\phi \neq 0$) are formulated. The functional derivatives of these equations with respect to $\phi$ are then computed. Finally, the source $\phi$ is set to zero to obtain the on-shell results.

A crucial step in the covariant scheme is the formulation of the off-shell equations ($GW$, $HGW$, or $SGW$). A generic binary operator, such as the charge density or spin, can be expressed in the form $X_{\vec k}(1) = \int d(\bar{2}\bar{3})\psi^*_{\vec k}(\bar{2}) K_X(1,\bar{2},\bar{3}) \psi_{\vec k}(\bar{3})$, where $K_X$ is a kernel specific to the operator $X$. By adding an external local source $\phi_{Y\vec k}(1)$ coupling to $Y$, the action becomes
\begin{equation}
    S[\psi^*,\psi;\phi]=S[\psi^*,\psi]-\sum_{\vec k}\int d(1)\phi_{Y\vec k}Y(1).
\end{equation}
The introduction of this source term is equivalent to shifting the non-interacting propagator:
\begin{equation}
    (G_{0\vec k})^{-1}(\bar1,\bar2;\phi)=(G_{0\vec k})^{-1}(\bar1,\bar2)+\int d(3)\phi_{Y\vec k}(3)K_Y(3,\bar1,\bar2).\label{eq:shift}
\end{equation}
The expectation value of the operator $X$ in the presence of the source is then given by
\begin{equation}
    \left< X_{\vec k}(1) \right>=\int d(\bar 2\bar 3)G_{\vec k}(\bar3,\bar2)K_X(1,\bar2,\bar3).
\end{equation}
The two-body correlation function is then obtained as
\begin{align}
    \chi_{XY,\vec k}(1,2)=&\frac{\delta \left< X_{\vec k}(1) \right>}{\delta \phi_{Y\vec k}(2)}=\int d(\bar 3\bar 4)\dot G_{\vec k}(\bar4,\bar3,2)K_X(1,\bar3,\bar4)\nonumber\\
    =&-\int d(\bar 3\bar 4\bar5\bar6)G_{\vec k}(\bar4,\bar5)\Gamma_{\vec k}(\bar5,\bar6,2)G_{\vec k}(\bar6,\bar3)K_X(1,\bar3,\bar4),
\end{align}
where we have introduced the covariant vertices:
\begin{equation}
        \dot G_{\vec k}(\bar1,\bar2,3)=\frac{\delta G_{\vec k}(\bar1,\bar2)}{\delta\phi_{Y\vec k}(3)},\label{eq:def-dotG}
\end{equation}
\begin{equation}
        \Gamma_{\vec k}(\bar1,\bar2,3)=\frac{\delta (G_{\vec k})^{-1}(\bar1,\bar2)}{\delta\phi_{Y\vec k}(3)}.\label{eq:def-Gamma}
\end{equation}
These two vertices are related by:
\begin{equation}
    \dot G_{\vec k}(\bar1,\bar2,3)=-\int d(\bar4\bar5)G_{\vec k}(\bar1,\bar4)\Gamma_{\vec k}(\bar4,\bar5,3)G_{\vec k}(\bar5,\bar2).
\end{equation}
For the charge correlation, the operator takes the form $X=Y=n$, and static charge susceptibility can be obtained through:
\begin{equation}
    \chi_c=\frac{\partial n}{\partial\mu}=\sum_{\vec k\sigma}\int d2\frac{\delta G_{\vec k\sigma}(1,1)}{\delta \phi_{c\vec k}(2)}=\sum_{\vec k}\chi_{c,\vec k}(i\omega_n=0).
\end{equation}
Similarly, the spin susceptibility should be calculated by setting $X=Y=S^z$ with:
\begin{equation}
    \chi_s=\frac{\partial m}{\partial h}=\sum_{\vec k\sigma}\int d2\sigma\frac{\delta {G}_{\vec k\sigma}( 1, 1)}{\delta \phi_{c\vec k}(2)}=\sum_{\vec k}\chi_{s,\vec k}(i\omega_n=0).
\end{equation}

Here, $\phi_{\vec kc}$ and $\phi_{\vec ks}$ denote the sources coupled to the charge density $n$ and spin $S^z$, respectively, used for calculating the covariant responses. For convenience, we define the covariant correlation function in momentum space as
\begin{subequations}
\begin{align}
    \chi_{c,\vec k}(1,2)=&\sum_{\sigma}\frac{\delta G_{\vec k\sigma}(1,1)}{\delta \phi_{c\vec k}(2)},\label{eq:covariant-chick}\\
    \chi_{s,\vec k}(1,2)=&\sum_{\sigma}\sigma\frac{\delta G_{\vec k\sigma}(1,1)}{\delta \phi_{s\vec k}(2)}.\label{eq:covariant-chisk}   
\end{align}
\end{subequations}
\subsubsection{covariant $GW$}

We now derive the covariant $GW$ (cGW) equations. Starting from the off-shell $GW$ equations, i.e., Eqs.~(\ref{eq:GW-Dyson}-\ref{eq:GW-H},\ref{eq:GW-Sigma},\ref{eq:eq:GW-P}), in which the non-interacting propagator is replaced by $(G_{0\vec k})^{-1}(\bar{1},\bar{2};\phi)$ according to Eq.~\eqref{eq:shift}, we take the functional derivative with respect to the source field $\phi_{Y\vec k}(3)$. This yields the set of covariant $GW$ (cGW) equations:
\begin{align}
    &\Gamma_{\vec k}(\bar1,\bar2,3)=\gamma_{\vec k}(\bar1,\bar2,3)+\Gamma_{\mathrm{H}\vec k}(\bar1,\bar2,3)+\Gamma_{\mathrm{MT}\vec k}(\bar1,\bar2,3)+\Gamma_{\mathrm{AL}\vec k}(\bar1,\bar2,3),\nonumber\\
    &\gamma_{\vec k}(\bar1,\bar2,3)=K_Y(3,\bar1,\bar2),\nonumber\\
    &\Gamma_{\mathrm{H}\vec k}(\bar1,\bar2,3)=-\delta(\bar1,\bar2)\int d\bar4 V(\bar1,\bar4)\dot{G}_{\vec k}(\bar4,\bar4,3),\nonumber\\
    &\Gamma_{\mathrm{MT}\vec k}(\bar1,\bar2,3)=\dot{G}_{\vec k}(\bar1,\bar2,3)W_{\vec k}(\bar2,\bar1),\nonumber\\
    &\Gamma_{\mathrm{AL}\vec k}(\bar1,\bar2,3)={G}_{\vec k}(\bar1,\bar2)\dot W_{\vec k}(\bar2,\bar1,3),\nonumber\\
    &\Gamma_{\mathrm{W}\vec k}(\bar1,\bar2,3)=-\dot G_{\vec k}(\bar1,\bar2,3)G_{\vec k}(\bar2,\bar1)-G_{\vec k}(\bar1,\bar2)\dot G_{\vec k}(\bar2,\bar1,3).\label{eq:cGW}
\end{align}
Here, we define the bosonic vertex
\begin{equation}
    \dot W_{\vec k}(\bar1,\bar2,3)=\frac{\delta W_{\vec k}(\bar1,\bar2)}{\delta\phi_{Y\vec k}(3)},
\end{equation}
\begin{equation}
    \Gamma_{\mathrm{W}\vec k}(\bar1,\bar2,3)=\frac{\delta W_{\vec k}^{-1}(\bar1,\bar2)}{\delta\phi_{Y\vec k}(3)},
\end{equation}
with the relation $\dot W_{\vec k}(\bar1,\bar2,3)=-\int d(\bar4\bar5)W_{\vec k}(\bar1,\bar4)\Gamma_{\mathrm{W}\vec k}(\bar4,\bar5,3)W_{\vec k}(\bar5,\bar2)$. The Eqs.(\ref{eq:cGW}) consist of the cGW equations, which are linear in the vertex $\Gamma$ and can be solved self-consistently. Once the vertex is determined, the charge and spin susceptibilities, $\chi_c$ and $\chi_s$, are computed via Eqs. \eqref{eq:covariant-chick} and \eqref{eq:covariant-chisk}. We note that the conventional random-phase approximation (RPA) within the $GW$ framework is recovered by retaining only the first two vertex terms: $\Gamma_{\mathrm{RPA}} = \gamma + \Gamma_{\mathrm{H}}$.

\subsubsection{covariant $HGW$}
The covariant $HGW$ (cHGW) equations are derived similarly by functional differentiation of the off-shell $HGW$ equations, i.e., Eqs.~(\ref{eq:HGW-Dyson}-\ref{eq:HGW-H}), following a procedure analogous to that used for cGW. This yields the cHGW equations:
\begin{align}
    &\Gamma_{\vec k}(\bar1,\bar2,3)=\gamma_{\vec k}(\bar1,\bar2,3)+\Gamma_{\mathrm{H}\vec k}(\bar1,\bar2,3)+\Gamma_{\mathrm{MT}\vec k}(\bar1,\bar2,3)+\Gamma_{\mathrm{AL}\vec k}(\bar1,\bar2,3),\nonumber\\
    &\gamma_{\vec k}(\bar1,\bar2,3)=K_Y(3,\bar1,\bar2),\nonumber\\
    &\Gamma_{\mathrm{H}\vec k}(\bar1,\bar2,3)=-\delta(\bar1,\bar2)\int d\bar4 V(\bar1,\bar4)\dot{G}_{\vec k}(\bar4,\bar4,3),\nonumber\\
    &\Gamma_{\mathrm{MT}\vec k}(\bar1,\bar2,3)=\dot{H}_{\vec k}(\bar1,\bar2,3)W_{\vec k}(\bar2,\bar1),\nonumber\\
    &\Gamma_{\mathrm{AL}\vec k}(\bar1,\bar2,3)={H}_{\vec k}(\bar1,\bar2)\dot W_{\vec k}(\bar2,\bar1,3),\nonumber\\
    &\Gamma_{\mathrm{W}\vec k}(\bar1,\bar2,3)=-\dot H_{\vec k}(\bar1,\bar2,3)G_{\vec k}(\bar2,\bar1)-H_{\vec k}(\bar1,\bar2)\dot G_{\vec k}(\bar2,\bar1,3).\label{eq:cHGW}
\end{align}
The vertex $\dot H$ can be calculated by
\begin{equation}
        \dot H_{\vec k}(\bar1,\bar2,3)=-\int d(\bar4\bar5)H_{\vec k}(\bar1,\bar4)[\gamma_{\vec k}(\bar4,\bar5,3)+\Gamma_{\mathrm{H}\vec k}(\bar4,\bar5,3)]H_{\vec k}(\bar5,\bar2).\label{eq:dotH}
\end{equation}
Together, Eqs.~\eqref{eq:cHGW} and Eq.~(\ref{eq:dotH}) form the self-consistent cHGW equations for the vertex, from which the corresponding susceptibilities are subsequently calculated.

\subsubsection{covariant $SGW$} 

The covariant formalism for the $SGW$ approximation (cSGW) follows a similar path, but requires careful treatment of the spin structure. Due to this structure, it is natural to express the vertices as matrices in spin space. We define $\underline{\Gamma}(1,2,3) \equiv \Gamma(\bar{1},\bar{2},3)$ and $\underline{\dot{G}}(1,2,3) \equiv \dot{G}(\bar{1},\bar{2},3)$ as these spin-space representations. The cSGW equations are then given by:
\begin{align}
&\underline{\Gamma}_{\vec k}(1,2,3)=\underline{\gamma}_{\vec k}(1,2,3)+\underline{\Gamma}_{\mathrm{H}\vec k}(1,2,3)+\underline{\Gamma}_{\mathrm{MT}\vec k}(1,2,3)+\underline{\Gamma}_{\mathrm{AL}\vec k}(1,2,3),\nonumber\\
&\underline{\gamma}_{\vec k}=K_Y(3,\bar1,\bar2),\nonumber\\
&\underline{\Gamma}_{\mathrm{H}\vec k}(1,2,3)=\delta(1,2)\sum_{ab}\int d(4)\underline{\tau}^aV^{ab}(1,4)\mathrm{tr}[\underline{\dot G}_{\vec k}(4,4,3)\underline{\tau}^b],\nonumber\\
&\underline{\Gamma}_{\mathrm{MT}\vec k}(1,2,3)=\sum_{ab}\underline{\tau}^a\underline{\dot G}_{\vec k}(1,2,3)\underline\tau^bW^{ba}_{\vec k}(2,1),\nonumber\\
&\underline{\Gamma}_{\mathrm{AL}\vec k}(1,2,3)=\sum_{ab}\underline{\tau}^a\underline{G}_{\vec k}(1,2)\underline\tau^b\dot W^{ba}_{\vec k}(2,1,3),\nonumber\\
&\Gamma_{\mathrm{W}}^{ab}(1,2,3)=-\mathrm{Tr}[\underline{\tau}^a\underline{\dot G}_{\vec k}(1,2,3)\underline{\tau}^b\underline{G}_{\vec k}(2,1)]-\mathrm{Tr}[\underline{\tau}^a\underline{G}_{\vec k}(1,2)\underline{\tau}^b\underline{\dot G}_{\vec k}(2,1,3)].\label{eq:cSGW}
\end{align}
Here, we define the bosonic vertex
\begin{equation}
    \Gamma_{\mathrm{W}\vec k}^{ab}(1,2,3)=\frac{\delta (W_{\vec k}^{-1})^{ab}(1,2)}{\delta\phi_{Y\vec k}(3)},
\end{equation}
\begin{equation}
    \dot W_{\vec k}^{ab}(1,2,3)=\frac{\delta W_{\vec k}^{ab}(1,2)}{\delta\phi_{Y\vec k}(3)}=-\sum_{cd}\int d(45)W^{ac}_{\vec k}(1,4)\Gamma^{cd}_{\mathrm{W}\vec k}(4,5,3)W^{db}_{\vec k}(5,2).
\end{equation}
These equations form the self-consistent cSGW equations, from which the spin and charge susceptibilities can be directly obtained via Eqs. \eqref{eq:covariant-chick} and \eqref{eq:covariant-chisk}.

The covariant framework establishes a fundamental standard for deriving response functions within many-body approximations like $GW$, $HGW$, and $SGW$.  It ensures the mathematical uniqueness of the susceptibilities and guarantees the preservation of fundamental identities, thereby resolving the ambiguities inherent in non-covariant treatments.

\section{Numerical Results}\label{Sec:Numerical}
In the previous section, we introduced the exact formula for the Green's function and the susceptibility, and established three different many-body approximate theories. Now we will present the single particle properties and the susceptibility results on the $64\times 64$ square lattice.

\subsection{Single Particle Properties}
\begin{figure}
 \centering
        \includegraphics[width=0.7\textwidth]{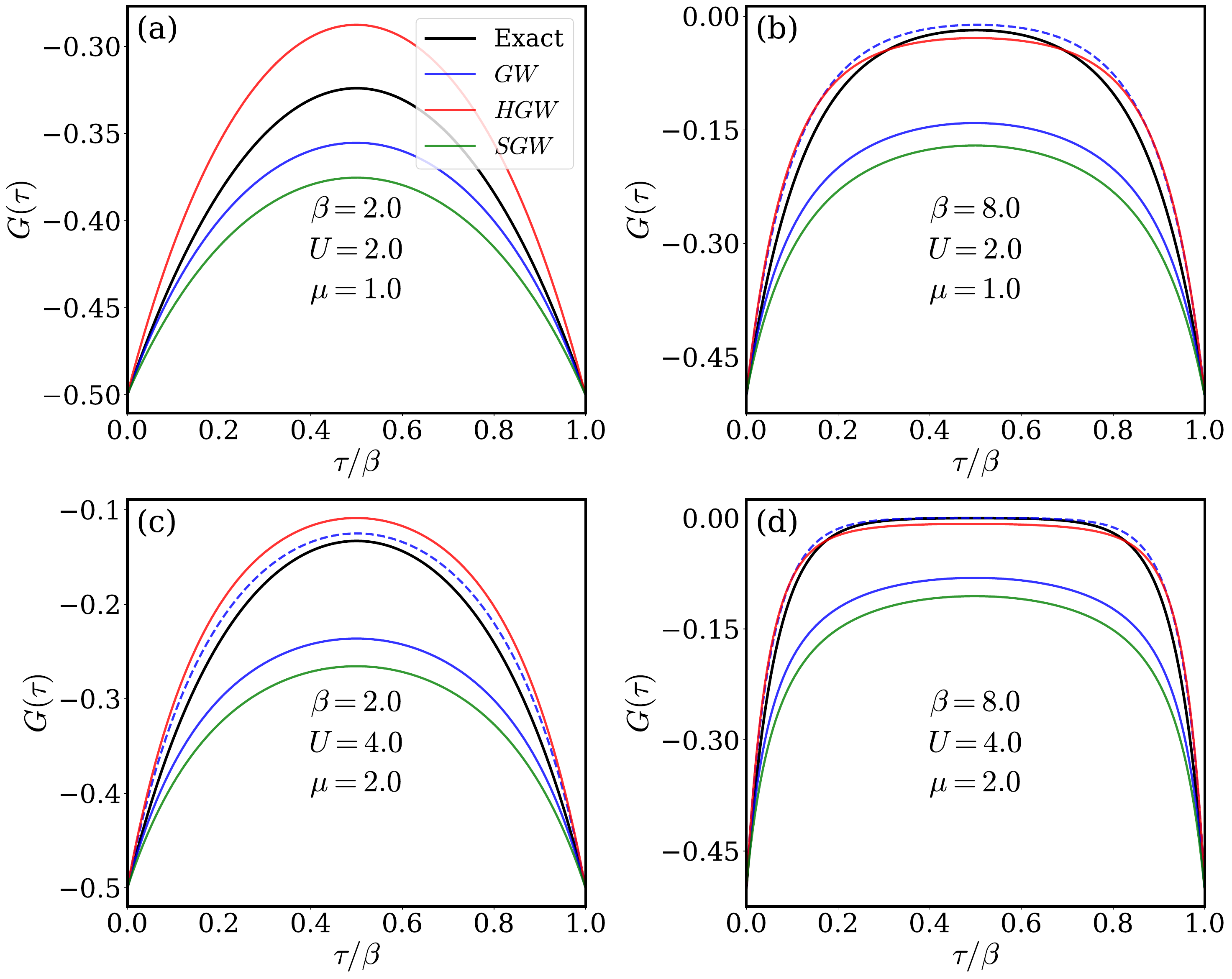}
 \caption{Comparison of imaginary-time Green's functions at half-filling for the antinodal point $k_{\mathrm{AN}}=(\pi,0)$. Results are shown for the exact solution (black lines), $GW$ (blue), $HGW$ (red), and $SGW$ (green) methods at different parameters: (a) $\beta=2, U=2$; (b) $\beta=8, U=2$; (c) $\beta=2, U=4$; (d) $\beta=8, U=4$. Different line styles for a given color correspond to different branches calculated by the same method.
}
\label{fig:Green-half}
\end{figure}

We begin by comparing the imaginary-time Green's functions $G(\vec{k}, \tau)$ at the antinodal point $\vec{k}_{\mathrm{AN}}=(\pi,0)$, computed using the exact solution and the many-body approximations. The comparison is performed for a set of temperatures ($\beta = 2, 8$) and interaction strengths ($U = 2, 4$). For many-body integral equation methods such as $GW$, $HGW$, and $SGW$, multiple solution branches can arise in certain parameter regions. It has conventionally been assumed that only one branch is physically meaningful, while the others often exhibit unphysical behavior—for instance, yielding non-real particle number densities. Moreover, it was commonly believed that $GW$ performs poorly at strong interaction strengths and low temperatures, failing to capture key phenomena such as the pseudo-gap and Mott insulating phases, whereas $HGW$ performs well in such regimes. However, in the HK model (Fig.~\ref{fig:Green-half}), the $GW$ method exhibits an additional solution branch at strong interactions and low temperatures that agrees closely with the exact solution, showing notably accurate results at low temperatures. In contrast, the $SGW$ approximation consistently shows significant deviations in the computed Green’s functions.

\begin{figure}
 \centering
        \includegraphics[width=0.7\textwidth]{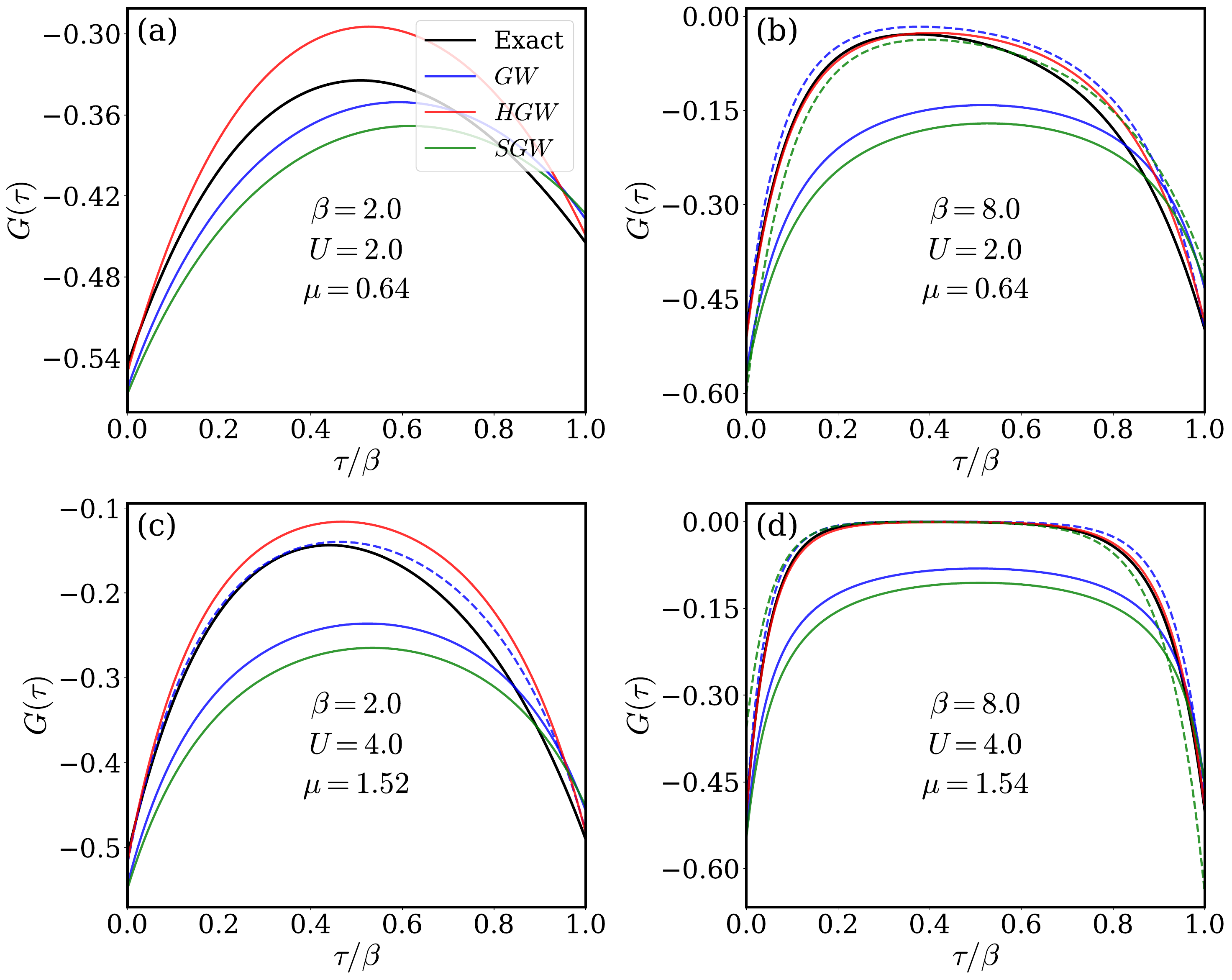}
 \caption{Imaginary-time Green's functions at the antinodal point $k_{\mathrm{AN}}=(\pi,0)$ for finite doping, comparing the exact solution (black lines) with the $GW$ (blue), $HGW$ (red), and $SGW$ (green) methods under different parameters: (a) $\beta=2, U=2$; (b) $\beta=8, U=2$; (c) $\beta=2, U=4$; (d) $\beta=8, U=4$. Different line styles within the same color represent distinct branches from the same method.
}
\label{fig:Green-dopping}
\end{figure}

To demonstrate the general applicability of many-body approaches away from half-filling, we compare the Green's functions obtained from different methods under finite doping in Fig.~(\ref{fig:Green-dopping}). Here, the chemical potential is fixed such that the exact particle density satisfies $n_{\vec k_{\mathrm{AN}}}=0.9$. As illustrated in Fig.~(\ref{fig:Green-dopping}), at finite doping with either strong interaction or low temperature, the $HGW$ method consistently yields accurate results. Both $GW$ and $SGW$ methods exhibit two branches: one produces unsatisfactory outcomes, while the other gives curves that closely align with the exact solution.

A key difference from the Hubbard model is that the exact solution of the HK model exhibits Mott insulating behavior even at weak couplings or high temperatures. This inherent property of the HK model explains why many-body approaches other than $HGW$ often fail in this regime, as the $GW$ approximation tends to predict metallic states in such a parameter region.

\begin{figure}
 \centering
        \includegraphics[width=0.7\textwidth]{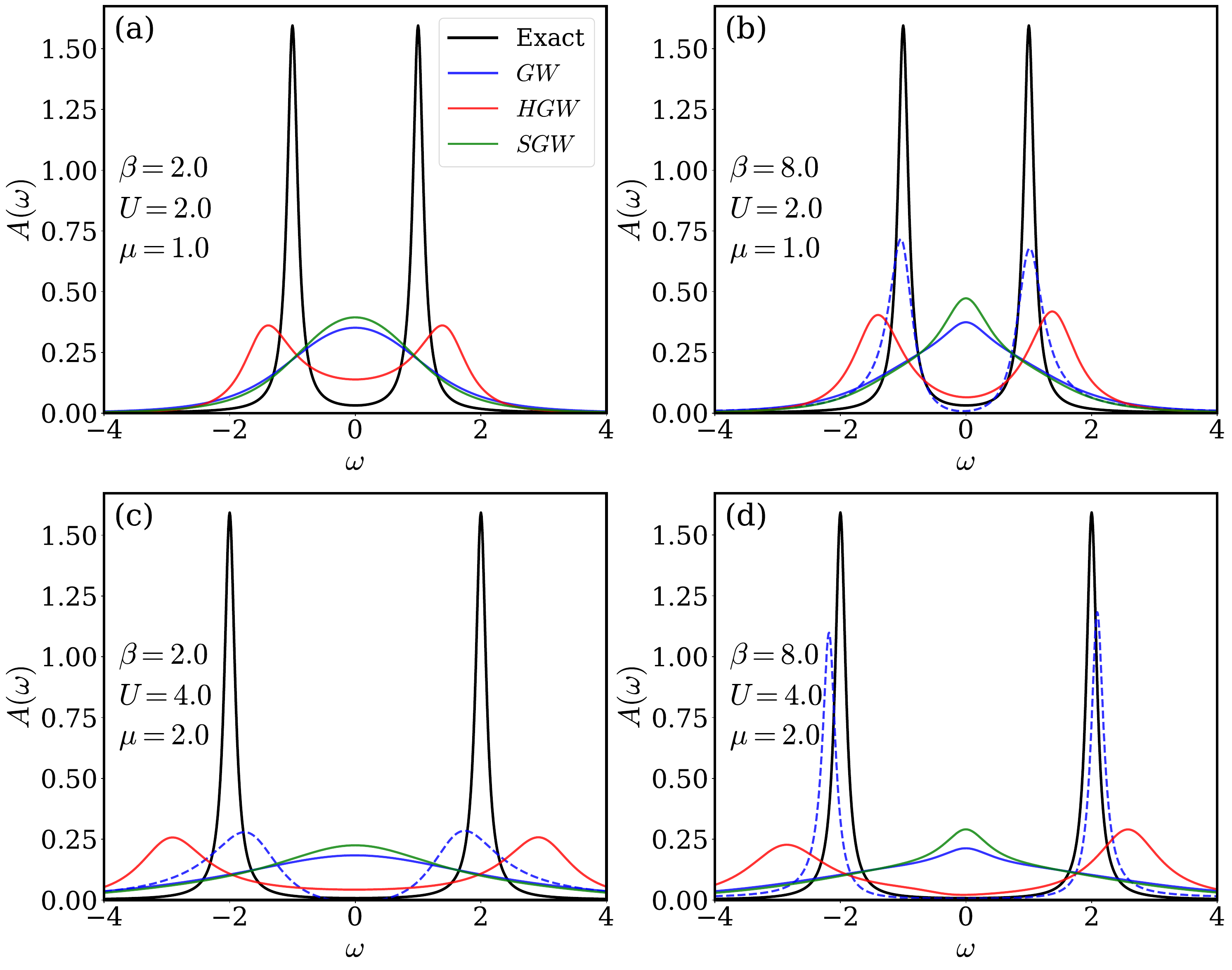}s
 \caption{Spectral functions at the antinodal point $k_{\mathrm{AN}}=(\pi,0)$ under half-filling are shown for the exact solution, $GW$, $HGW$, and $SGW$ methods across different parameter sets: (a) $\beta=2, U=2$; (b) $\beta=8, U=2$; (c) $\beta=2, U=4$; (d) $\beta=8, U=4$. The black, blue, red, and green lines represent results from the exact formula, $GW$, $HGW$, and $SGW$ approximations, respectively. Notably, the $GW$ method yields a two-peak Mott insulating structure.}
\label{fig:specral_half_filling}
\end{figure}

\begin{figure}
 \centering
        \includegraphics[width=0.7\textwidth]{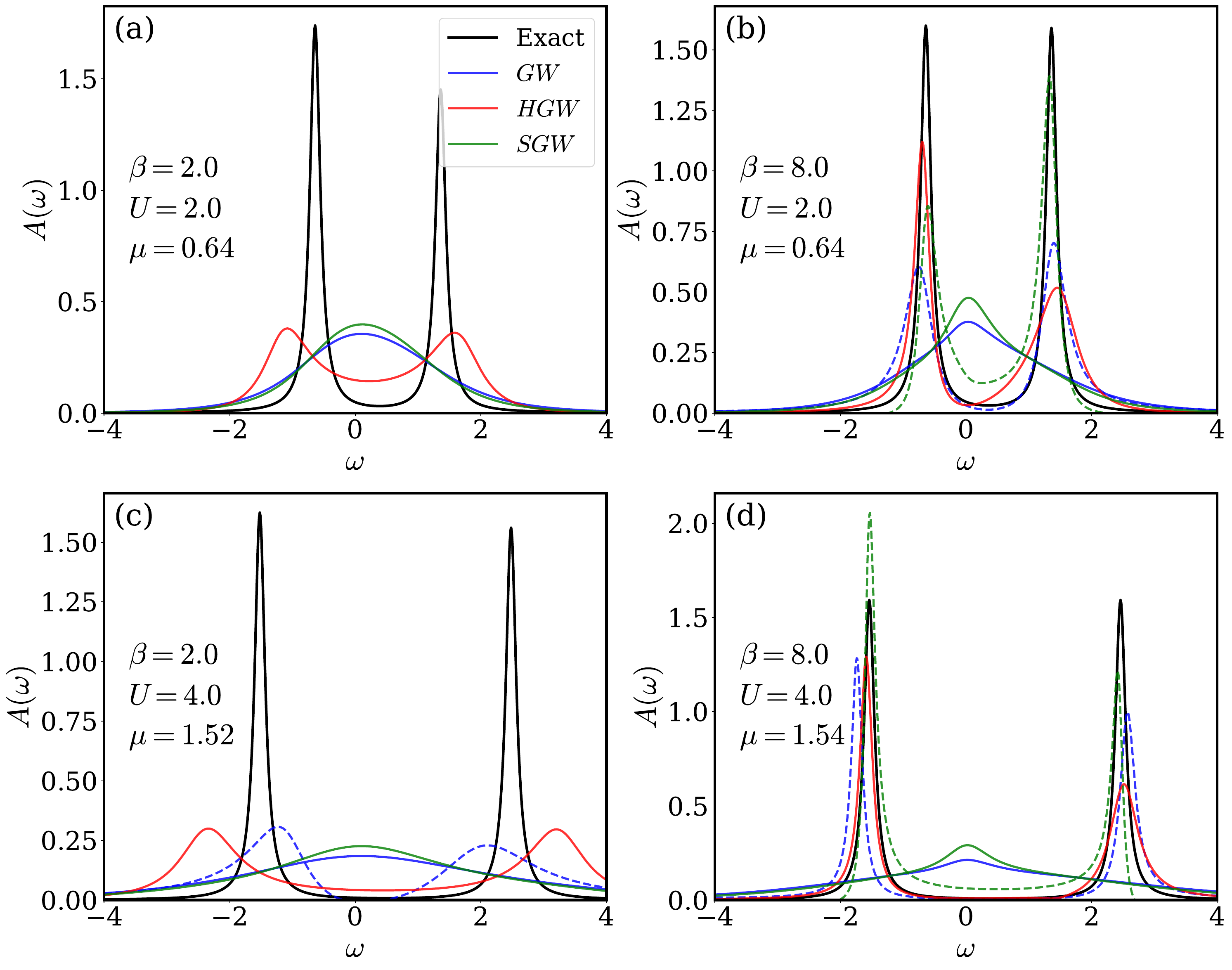}
 \caption{Spectral functions at the antinodal point $k_{\mathrm{AN}}=(\pi,0)$ under finite doping, comparing the exact solution (black lines) with the $GW$ (blue), $HGW$ (red), and $SGW$ (green) methods. Results are shown for different parameter sets: (a) $\beta=2, U=2$; (b) $\beta=8, U=2$; (c) $\beta=2, U=4$; (d) $\beta=8, U=4$. Both the $GW$ and $SGW$ methods exhibit an additional spectral branch, which manifests as an extra peak associated with the Mott structure.}
\label{fig:specral_dopping}
\end{figure}

To provide deeper insight into the single-particle excitation properties, we systematically computed the spectral function \( A(\vec{k},\omega) \), which was obtained by performing the analytical continuation of the Green's function using the AAA algorithm\cite{aaa}. The key results under both half-filling and finite doping are presented in Figs.~\ref{fig:specral_half_filling} and~\ref{fig:specral_dopping}, respectively. A central finding of our spectral analysis is that the conventional $GW$ approximation can yield a Mott insulating solution in this model whose gap width is in excellent quantitative agreement with the exact result. Specifically, at half-filling (Fig.~\ref{fig:specral_half_filling}), under lower temperatures (e.g., \( \beta = 8 \)), the $GW$ method produces—in addition to a common single-peak metallic solution—a second branch exhibiting a clear Mott gap, whose width closely matches that of the exact solution. This behavior remains stable for both \( U = 2 \) and \( U = 4 \). In comparison, while the $HGW$ method consistently produces a Mott solution, it systematically overestimates the gap. The $SGW$ method, in most cases, only yields a single-peak solution. 

In the finite doping case (Fig.~\ref{fig:specral_dopping}), the performance of the $GW$ approach is further validated: at low temperature, $SGW$ also develops a branch with an accurate Mott gap. These results collectively demonstrate that $GW$ and $SGW$ are capable of capturing the Mott insulating phase in this model—a finding that challenges the conventional understanding that $GW$ cannot describe strongly correlated Mott physics. It should be noted that all many-body methods considered produce spectral functions with asymmetric peak heights differing from the exact results, pointing to a direction for future theoretical refinement.

\begin{figure}
 \centering
        \includegraphics[width=0.8\textwidth]{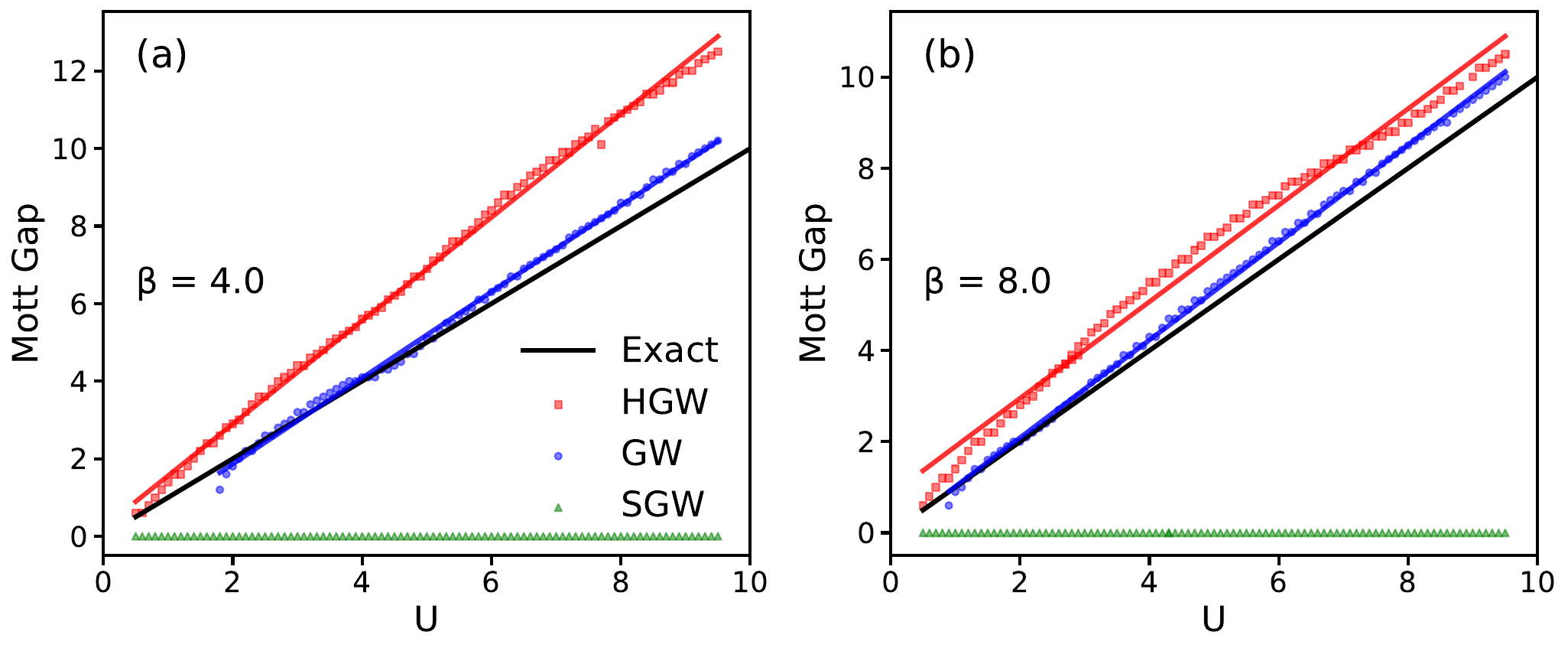}
 \caption{Mott gap as a function of interaction strength $U$ at the antinodal point $k=(\pi,0)$ under half-filling. Panels (a) and (b) show results for $\beta=4$ and $\beta=8$, respectively. The exact solution is represented by a black line, while results from the $GW$ (blue), $SGW$ (green), and $HGW$ (red) methods are shown as discrete points. The $GW$ and $HGW$ data points are fitted with linear functions (dashed lines). For $\beta=4$ (a), the fitted slope is 1.11 ($R^2=0.998$) for $GW$ and 1.33 ($R^2=0.998$) for $HGW$. For $\beta=8$ (b), the fitted slope is 1.07 ($R^2=0.999$) for $GW$ and 1.06 ($R^2=0.987$) for $HGW$.}
\label{fig:mott_gap}
\end{figure}

We further quantify the Mott gap $\Delta_{\mathrm{Mott}}$ as a function of interaction strength $U$ at half-filling (Fig.~\ref{fig:mott_gap}). The exact solution of the spectral function manifests as a linear relation between the gap and interaction, $\Delta_{\mathrm{Mott}} = U$. The $GW$ method captures this linear dependence exceptionally well across temperatures ($\beta=4, 8$), with fitted slopes of $A_{GW} \approx 1.07$–$1.11$ and coefficients of determination $R^2 > 0.998$. While the $HGW$ method also produces a Mott gap that increases with $U$, its linearity is poorer ($R^2 \approx 0.987$ at $\beta=8$) and it systematically overestimates the gap magnitude across nearly the entire parameter range. This quantitative comparison highlights the unexpected capability of the $GW$ approximation in describing the Mott transition within the HK model.

\subsection{Two-body Properties}
\begin{figure}
 \centering
        \includegraphics[width=0.8\textwidth]{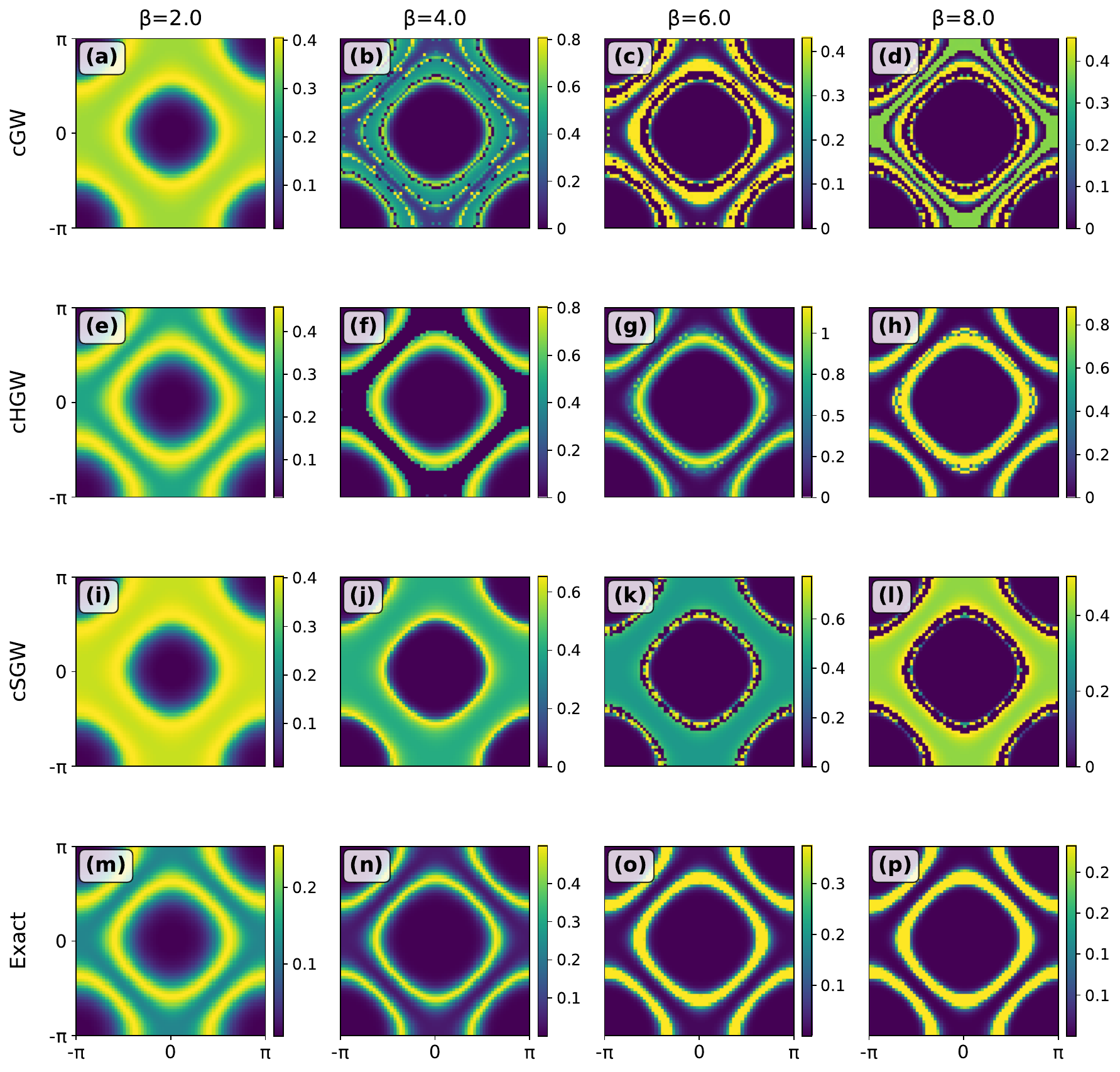}
 \caption{Charge susceptibility in the first Brillouin zone for a half-filled $64\times64$ system. The panels are organized by method (rows) and inverse temperature $\beta$ (columns): rows 1 to 4 correspond to the cGW, cHGW, cSGW, and Exact methods, respectively; columns 1 to 4 correspond to $\beta=2,4,6,8$, respectively.}
\label{fig:charge_k}
\end{figure}

\begin{figure}
 \centering
        \includegraphics[width=0.8\textwidth]{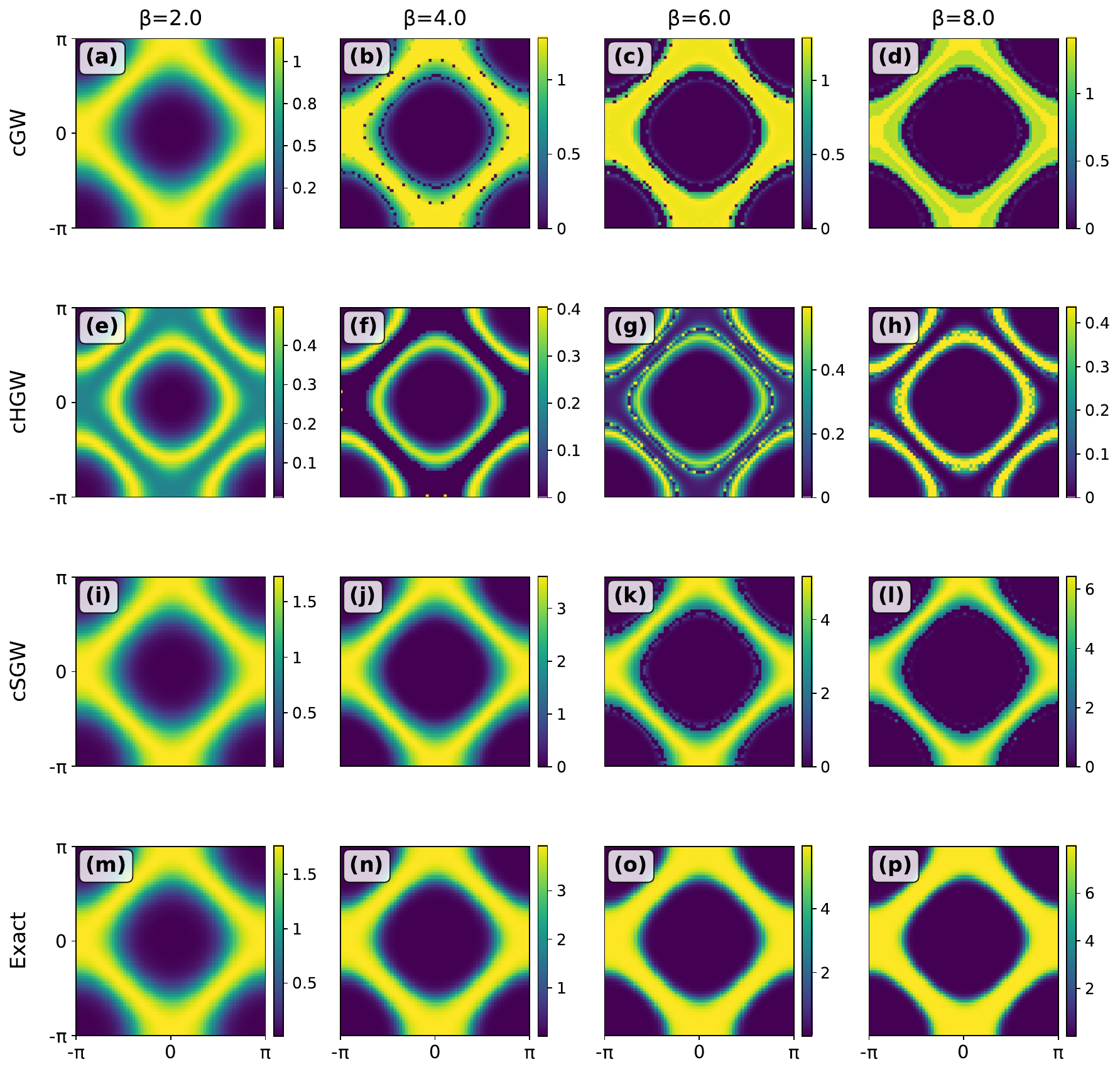}
 \caption{Spin susceptibility in the first Brillouin zone for a half-filled $64\times64$ system. The 4×4 panel array compares the cGW, cHGW, cSGW, and Exact methods (rows 1-4) across temperatures corresponding to $\beta=2,4,6,8$ (columns 1-4).}
\label{fig:spin_k}
\end{figure}

In this part, we study the response properties of $GW$, $HGW$, and $SGW$ approaches in the HK model. For the HK model, the Hilbert space is decoupled in the momentum space, therefore, the spin or the charge response is also the consist of different momentum components independently, which can be written as
\begin{equation}
    \chi_c=\frac{1}{L^2}\sum_{\vec k}\chi_c(\vec k),\quad \chi_s=\frac{1}{L^2}\sum_{\vec k}\chi_s(\vec k).
\end{equation}
For the exact solution, the momentum-dependent response can be simulated through:
\begin{align}
    \chi_c(\vec k)=&\sum_\sigma\frac{\partial n_{\vec k\sigma}}{\partial \mu},\nonumber\\
    \chi_s(\vec k)=&\frac{\partial}{\partial h}(n_{\vec k\uparrow}-n_{\vec k\downarrow})\left.\right|_{h\rightarrow0}.
\end{align}
As for the covariant framework for different many-body approximate theory, the momentum-dependent responses $\chi_c(\vec k)$ and $\chi_s(\vec k)$ can be simulated through Eqs.~(\ref{eq:covariant-chick},\ref{eq:covariant-chisk}) directly. To study the momentum structure of the response function in the covariant framework, we plot the momentum-dependent charge and spin susceptibility at half-filling for $U=2$ with different temperatures in Fig.~\ref{fig:charge_k} and Fig.~\ref{fig:spin_k}. It should be pointed out that, the integral equation has a multi-solution problem, and in some parameters, the physically stable solution is very difficult to obtain, which would lead to the discontinuity in Fig.~\ref{fig:charge_k} and Fig.~\ref{fig:spin_k}. However, the global distribution structure of the response is independent of the instability of the specific momentum solution.

Our benchmark of the charge response in momentum space reveals a unique structure, as shown in Fig.~\ref{fig:charge_k}. The response is strongly suppressed on the non-interacting Fermi surface at half-filling, given by $k_x \pm k_y = ±\pi$. Significant non-zero contributions instead form a closed loop inside this surface and symmetric arcs outside it. This characteristic pattern becomes increasingly pronounced as the temperature decreases. Among the methods tested, only cHGW correctly captures both this spatial distribution and its temperature dependence. Crucially, cHGW reproduces the near-zero response on the Fermi surface at low temperatures. In contrast, both cGW and cSGW perform poorly for this charge property, as they fail to show the suppressed response on this key line. 

The momentum distribution of the spin response presents a strikingly different picture from the charge, as shown in Fig.~\ref{fig:spin_k}. It shows significant strength on and around a broad region of the non-interacting Fermi surface $k_x \pm k_y = ±\pi$, forming a large, connected area of enhanced response. For this phenomenon, both cGW and cSGW can qualitatively capture the basic shape of the distribution. However, they differ greatly in quantitative accuracy: cSGW results agree well with the exact solution in the magnitude of the response, while cGW significantly underestimates the strength of the spin correlations. This is not an isolated finding; we have observed similar results in the Hubbard model, where cSGW accurately computes spin correlations and cHGW excels at charge correlations. This consistency strengthens our understanding of these methods' applicability.

\section{Conclusion and Discussion}\label{Sec:Conclusion}

In summary, we systematically apply three distinct many-body approximation theories, $GW$, $HGW$, and $SGW$, along with a covariant framework for computing two-particle correlation functions, to the HK model. For single-particle properties, we compare the imaginary-time Green's functions obtained from the three approximate methods with the exact solution across different temperatures, interaction strengths, and at both half-filling and non-half-filling. We find that $HGW$ consistently yields reasonably good results, while $GW$, under conditions of low temperature or large U, generates a solution branch that closely matches the exact result. This finding challenges the conventional understanding that $GW$ performs poorly in strongly correlated regimes.

To further investigate the capabilities of these methods in describing Mott physics, we employ the AAA algorithm\cite{aaa} for analytic continuation of the Green's functions obtained from each method to extract the single-particle spectral function. Results show that $HGW$ always produces a gap, but the predicted gap width is typically significantly wider than the exact gap, consistent with its behavior in the Hubbard model. Surprisingly, in this model, the $GW$ theory successfully yields a Mott solution at low temperatures or large U, and the associated gap shows remarkable agreement with the exact gap. Moreover, the relationship between the Mott gap obtained from $GW$ and the interaction strength U closely follows the exact linear relation gap = U, with a deviation in the slope of only $7\%$, and the data fits a linear relationship with an R² value as high as 0.9991.

For the response functions, the spatial and temperature dependence of the charge response is accurately captured only by cHGW. In contrast, the strength of the spin response is well reproduced by cSGW but significantly underestimated by cGW. These trends align with observations in the Hubbard model, suggesting a consistent performance pattern across different correlated systems.

This study confirms the strengths of $HGW$ and $SGW$ in capturing charge and spin responses, respectively, and unexpectedly reveals the capability of $GW$ in describing Mott physics under specific conditions. By benchmarking against exact solutions, our results help clarify the applicability of these many-body approximations, providing useful guidance for their application in future studies of strongly correlated phenomena.

\begin{acknowledgments}
 This work is supported by MOST 2022YFA1402701 and the High-performance
 Computing Platform of Peking University. The authors are very grateful to Dingping Li for valuable discussions.
\end{acknowledgments}

\nocite{*}

\bibliography{ref}

\end{document}